\documentclass[11pt,a4paper]{article}

\usepackage[utf8]{inputenc}
\usepackage[T1]{fontenc}
\usepackage{lmodern}
\usepackage[margin=1in]{geometry}
\usepackage{amsmath,amssymb}
\usepackage{graphicx}
\usepackage{booktabs}
\usepackage{multirow}
\usepackage{enumitem}
\usepackage{xcolor}
\usepackage{listings}
\usepackage{tikz}
\usetikzlibrary{positioning, arrows.meta, shapes.geometric, fit, calc, backgrounds}
\usepackage[colorlinks=true,linkcolor=blue!60!black,citecolor=blue!60!black,urlcolor=blue!60!black]{hyperref}
\hypersetup{
  pdftitle={Agentic Education: Using Claude Code to Teach Claude Code},
  pdfauthor={Zain Naboulsi},
  pdfsubject={AI coding assistants, curriculum design, pedagogical agents},
  pdfkeywords={AI coding assistants, Claude Code, curriculum design, prompt engineering, auto-updating curricula, gradual release of responsibility, adaptive learning, agentic coding tools, self-efficacy}
}
\usepackage{natbib}
\usepackage{caption}
\usepackage{subcaption}
\usepackage{tabularx}
\usepackage{array}
\usepackage{float}
\usepackage{fontawesome5}
\usepackage[most]{tcolorbox}

\definecolor{sparqcoral}{HTML}{C94A2A}
\definecolor{sparqcorallight}{HTML}{FFF5F2}
\definecolor{sparqcoralborder}{HTML}{E8634A}

\definecolor{codebg}{rgb}{0.96,0.96,0.96}
\definecolor{codegreen}{rgb}{0.0,0.5,0.0}
\definecolor{codegray}{rgb}{0.5,0.5,0.5}
\definecolor{codepurple}{rgb}{0.58,0,0.82}

\lstdefinestyle{default}{
  backgroundcolor=\color{codebg},
  basicstyle=\ttfamily\small,
  breaklines=true,
  captionpos=b,
  commentstyle=\color{codegreen},
  keywordstyle=\color{codepurple}\bfseries,
  stringstyle=\color{red!60!black},
  numberstyle=\tiny\color{codegray},
  frame=single,
  framerule=0.4pt,
  rulecolor=\color{codegray!40},
  xleftmargin=1em,
  framexleftmargin=0.5em,
}
\definecolor{guidecolor}{RGB}{76,175,80}
\definecolor{collabcolor}{RGB}{33,150,243}
\definecolor{peercolor}{RGB}{255,152,0}
\definecolor{launchercolor}{RGB}{244,67,54}

\begin{document}

\begin{center}
  {\LARGE\bfseries\color{sparqcoral}%
    Agentic Education:\\[0.3em]
    Using Claude Code to Teach Claude Code\par}
  \vspace{1em}
  {\large Zain Naboulsi}\\[0.3em]
  Principal AI Engineer, Sparq\\[0.2em]
  \texttt{zain.naboulsi@teamsparq.com}
\end{center}

\vspace{0.8em}

\begin{tcolorbox}[
  colback=sparqcorallight,
  colframe=sparqcoralborder,
  arc=3pt,
  boxrule=1pt,
  left=10pt, right=10pt, top=10pt, bottom=10pt,
  title={\textbf{Abstract}},
  fonttitle=\normalsize,
  coltitle=white,
  colbacktitle=sparqcoral,
  attach boxed title to top left={yshift=-2mm, xshift=5mm},
  boxed title style={arc=2pt, boxrule=0pt}
]
AI coding assistants have proliferated rapidly, yet structured pedagogical frameworks for learning these tools remain scarce. Developers face a gap between tool documentation and practical mastery, relying on fragmented resources such as blog posts, video tutorials, and trial-and-error. We present \textsc{cc-self-train}, a modular interactive curriculum for learning Claude Code, an agentic AI coding tool, through hands-on project construction. The system introduces five contributions: (1)~a \emph{persona progression model} that adapts the AI instructor's tone and scaffolding across four stages (Guide $\to$ Collaborator $\to$ Peer $\to$ Launcher), operationalizing the Gradual Release of Responsibility framework for AI-mediated instruction; (2)~an \emph{adaptive learning system} that observes engagement quality through hook-based heuristics and adjusts scaffolding at two timescales: streak detection triggers mid-module intervention while aggregate metrics adjust persona schedules at module boundaries, grounded in evidence that engagement quality mediates AI-tutoring gains~\citep{chung2025adaptive} and that sequential failure patterns carry higher informational weight than isolated events~\citep{hooshyar2026neural}; (3)~a \emph{cross-domain unified curriculum} in which five distinct project domains share identical feature sequencing, enabling transfer learning; (4)~a \emph{step-pacing mechanism} with explicit pause primitives to manage information overload in an AI-as-instructor context; and (5)~an \emph{auto-updating curriculum design} in which the onboarding agent detects upstream tool changes and updates teaching materials before instruction begins. A parametrized test suite enforces structural consistency as a proxy for pedagogical invariants across all 50 modules. A pilot evaluation with 27 participants shows statistically significant reported self-efficacy gains across all 10 assessed skill areas ($p < 0.001$), with the largest effects on advanced features such as hooks and custom skills. We describe the system architecture, analyze each contribution through established pedagogical and AI agent design lenses, and discuss implications for the design of auto-updating educational systems.

\vspace{0.6em}
\noindent\begin{minipage}[b]{0.75\linewidth}
  \small\faGithub\ \url{https://github.com/zainnab-sparq/cc-self-train}
\end{minipage}%
\hfill
\begin{minipage}[b]{0.2\linewidth}
  \centering\raggedleft
  \includegraphics[height=1cm]{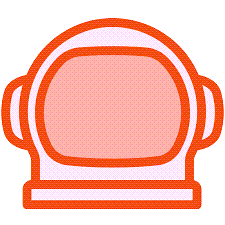}\\[-2pt]
  {\small\textsf{\textbf{Sparq}}}
\end{minipage}
\end{tcolorbox}

\vspace{0.4em}

\section{Introduction}
\label{sec:introduction}

The landscape of AI coding tools has expanded across several distinct categories in a short span. Early tools like Tabnine and Kite offered AI-powered code completion using smaller models, and GitHub Copilot~\citep{github2022copilot,chen2021evaluating} brought LLM-based completion to mainstream adoption. Cursor~\citep{cursor2024} extended this with codebase-aware context and chat-based interaction within a full IDE. More recently, \emph{agentic coding tools} such as Claude Code~\citep{anthropic2025claudecode} and OpenAI Codex~\citep{openai2025codex} introduced a different paradigm: they operate autonomously in terminal or cloud environments, planning multi-step tasks, executing shell commands, reading and writing files, and iterating on their own output~\citep{wang2024survey}. These categories are converging, with Copilot adding agent mode and Cursor gaining agentic capabilities, making the need for pedagogical frameworks for agentic tools increasingly broad. Developers face a widening gap between the availability of powerful AI agents and the learning pathways needed to use them effectively.

The cost of this gap is now visible at organizational scale. A 2025 industry survey by MIT's NANDA initiative reports that the overwhelming majority of enterprise generative-AI pilots fail to produce measurable operational or financial impact, locating the failure in a \emph{learning gap}: deployed systems neither retain feedback nor adapt to context, and users who succeed with AI in personal workflows often describe the same tools as unreliable inside enterprise systems~\citep{challapally2025genai}. The diagnosis admits two responses: build AI that learns from its users, or build learning environments that adapt users to the tool. This work pursues the second, focused on agentic coding tools.

The prevailing approach to learning AI coding tools is ad hoc and, critically, perishable. Official documentation describes features in isolation without progressive learning paths. Video tutorials and blog posts cover narrow use cases but date quickly. Agentic tools ship breaking changes on a rapid cadence measured in days. We observed nominally current third-party tutorials referencing deprecated or modified features within days of publication. Even vendor-produced training courses, including Anthropic's own course catalog~\citep{anthropic2025courses}, can lag behind the tool's shipping pace, leaving learners to reconcile instructional materials with a product that has already moved on. None of these resources provides a progressive, hands-on curriculum that takes a developer from first contact through advanced multi-agent orchestration, and none addresses the \emph{content decay} problem inherent in rapidly evolving tooling. This gap likely limits adoption of features that require compositional understanding, such as hook pipelines, skill systems, and subagent architectures.

We present \textsc{cc-self-train}, a curriculum framework designed to address this gap. While the current implementation targets Claude Code, a terminal-based agentic coding tool, the underlying pedagogical architecture (persona progression, unified curricula, step-pacing, and auto-updating design) is designed to generalize to any agentic environment, whether the tool is a coding assistant, a research agent, or a domain-specific AI workflow, and whether it operates in a terminal, an IDE, or the cloud. The curriculum currently comprises 50 module files organized as 10 progressive modules across 5 project domains. Rather than describing features abstractly, each module teaches Claude Code capabilities through the construction of a real software project, such as a portfolio website, a CLI toolkit, an API gateway, a code analyzer, or the learner's own existing project. While the curriculum provides sequenced scaffolding, controlling \emph{what} is taught and \emph{when}, the actual learning experience is exploratory and agent-driven. The learner is not clicking through slides or following a fixed script; they are conversing with an AI agent that responds to their specific project, questions, and mistakes, making each path through the material unique. The curriculum's pedagogical design draws on the author's 18 years of experience as a Microsoft Certified Trainer.

The curriculum is designed to stay current: when a learner begins, the onboarding agent checks whether its teaching materials match the learner's installed tool version and updates them if needed. This addresses the content decay problem at the point of use rather than relying on manual maintenance cycles.

\paragraph{Contributions.} This paper makes the following contributions:

\begin{enumerate}[leftmargin=2em]
  \item A \textbf{persona progression model} (Section~\ref{sec:persona}) that adapts instructor tone and scaffolding across four stages, mapping the Gradual Release of Responsibility (GRR) framework~\citep{fisher2013better} to AI-mediated instruction.

  \item An \textbf{adaptive learning system} (Section~\ref{sec:adaptive}) that observes learner engagement quality through hook-based heuristics and adjusts scaffolding at two timescales: streak detection for mid-module intervention and aggregate metrics for module-boundary persona changes. The design is grounded in evidence that engagement quality mediates AI-tutoring gains~\citep{chung2025adaptive} and that non-mastery signals carry higher gradient sensitivity than mastery signals in learner models~\citep{hooshyar2026neural}.

  \item A \textbf{cross-domain unified curriculum} (Section~\ref{sec:crossdomain}) that enforces identical feature sequencing across five project domains, designed to enable domain-independent transfer of tool mastery.

  \item A \textbf{step-pacing mechanism} (Section~\ref{sec:pacing}) with explicit pause primitives and cross-session state tracking, addressing the information overload problem characteristic of AI-as-instructor contexts.

  \item An \textbf{auto-updating curriculum design} (Section~\ref{sec:staying-current}) in which the onboarding agent detects upstream tool changes and updates teaching materials before instruction begins.
\end{enumerate}

\noindent A parametrized test suite (Section~\ref{sec:qa}) enforces structural consistency as a proxy for pedagogical invariants across all 50 modules, serving as the quality assurance layer for the above contributions. A pilot evaluation (Section~\ref{sec:evaluation}) with 27 participants provides initial empirical evidence that the curriculum produces significant reported self-efficacy gains across all assessed skill areas.

\section{Background and Related Work}
\label{sec:background}

\subsection{AI Coding Assistants}

AI coding tools exist on a spectrum of autonomy, and the boundaries between categories are blurring rapidly.

\paragraph{Completion and chat assistants.} Early AI code completion tools such as Tabnine and Kite used smaller models to offer inline suggestions. GitHub Copilot~\citep{github2022copilot,chen2021evaluating} brought LLM-based completion to mainstream adoption, and Cursor~\citep{cursor2024} extended this with codebase-aware context and chat-based interaction within a full IDE. In their original form, these tools augment the developer's existing workflow: the human remains the driver, and the AI provides suggestions within the editor.

\paragraph{Agentic coding tools.} Claude Code~\citep{anthropic2025claudecode} and OpenAI Codex~\citep{openai2025codex} shifted the developer's role from driver to reviewer. These tools operate autonomously, planning multi-step tasks, executing shell commands, reading and writing files, managing git workflows, and iterating on their own output. Claude Code exposes an extensible architecture of hooks, skills, subagents, MCP servers, and evaluation frameworks that compose into complex workflows.

\paragraph{Convergence.} These categories are converging: Copilot has added agent mode with autonomous multi-file editing and terminal access; Cursor has evolved from a chat-based assistant to a full agentic coding environment; and a growing number of IDEs, including Amazon Kiro and Google's Antigravity, are shipping agentic capabilities as core features. As agentic features become standard across tools, the need for pedagogical frameworks that teach compositional tool mastery grows correspondingly. An agentic system with dozens of interacting features cannot be learned through casual use alone, regardless of whether it runs in a terminal, an IDE, or the cloud. Our curriculum framework addresses this need; the current implementation targets Claude Code, but the pedagogical architecture is designed to be tool-agnostic.

\subsection{Developer Education for AI Tools}

Developer education for AI coding tools exists primarily in three forms: official documentation, informal tutorials, and structured courses. Official documentation usually describes features in isolation without progressive learning paths. YouTube tutorials and blog posts tend to cover narrow use cases (``How to use Copilot for X'') without building compositional understanding. Bootcamps and workshops often focus on prompt engineering~\citep{white2023prompt} rather than tool-specific mastery. \citet{becker2023programming} and \citet{kazemitabaar2023studying} study how AI code generators affect programming education, but focus on learner outcomes with code completion rather than curricula for tool mastery. \citet{prather2024widening} find that generative AI compounds existing metacognitive difficulties for struggling novice programmers, widening rather than closing the skill gap, motivating the need for structured learning pathways. Recent systems have begun embedding pedagogical guardrails directly into AI coding assistants: CodeAid~\citep{kazemitabaar2024codeaid} deploys an LLM-based assistant that provides conceptual explanations and pseudocode rather than direct solutions, demonstrating that pedagogical constraints can be designed into AI tool interactions, though the focus remains on teaching \emph{programming}, not on teaching mastery of an AI tool itself.

A recent scoping review of 52 studies characterizes the emerging field of LLM-based pedagogical agents along four design dimensions: interaction approach (reactive vs.\ proactive), domain scope (specific vs.\ general), role complexity (single vs.\ multi-role), and system integration (standalone vs.\ integrated)~\citep{li2026scoping}. Within this taxonomy, \textsc{cc-self-train} is a hybrid interaction agent (reactive to student-initiated code work, proactive through engagement observation and streak detection), domain-specific (targeting Claude Code mastery), multi-role (the persona transitions across four GRR stages), and integrated (hooking into the runtime it teaches). The reviewed literature covers agents that teach \emph{external} subjects: science, language, general programming concepts. Vendor-produced interactive onboarding touches on host-tool usage, but to our knowledge no prior LLM-based pedagogical agent adopts a staged-persona GRR curriculum for teaching its own host tool. This reflexive design, where the tool being learned is simultaneously the pedagogical medium, is the positioning gap \textsc{cc-self-train} fills.

A complementary line of work has emerged in the open-source community. Everything Claude Code~\citep{affaan2025ecc} provides a production-oriented configuration system comprising 16 specialized agents, 65+ skills, and 40+ commands that encode best practices for working with Claude Code effectively. OpenClaw~\citep{openclaw2025} takes a broader approach, offering an open-source personal AI assistant platform with extensible skills and multi-channel integration. Superpowers~\citep{obra2025superpowers} provides a structured workflow framework (brainstorming, planning, TDD, code review) that runs across multiple agentic tools, including Claude Code, Cursor, Codex, and Gemini CLI, demonstrating that structured agentic skill frameworks can be tool-agnostic. These projects collectively demonstrate the community demand for structured approaches to AI tool mastery and influenced the design of \textsc{cc-self-train}. Where Everything Claude Code and Superpowers optimize expert workflows and OpenClaw provides a general-purpose agent framework, our work focuses specifically on the \emph{pedagogical} dimension: progressive skill acquisition through a structured curriculum with adaptive instruction.

\subsection{Pedagogical Foundations}

\begin{figure}[t]
\centering
\includegraphics[width=0.75\textwidth]{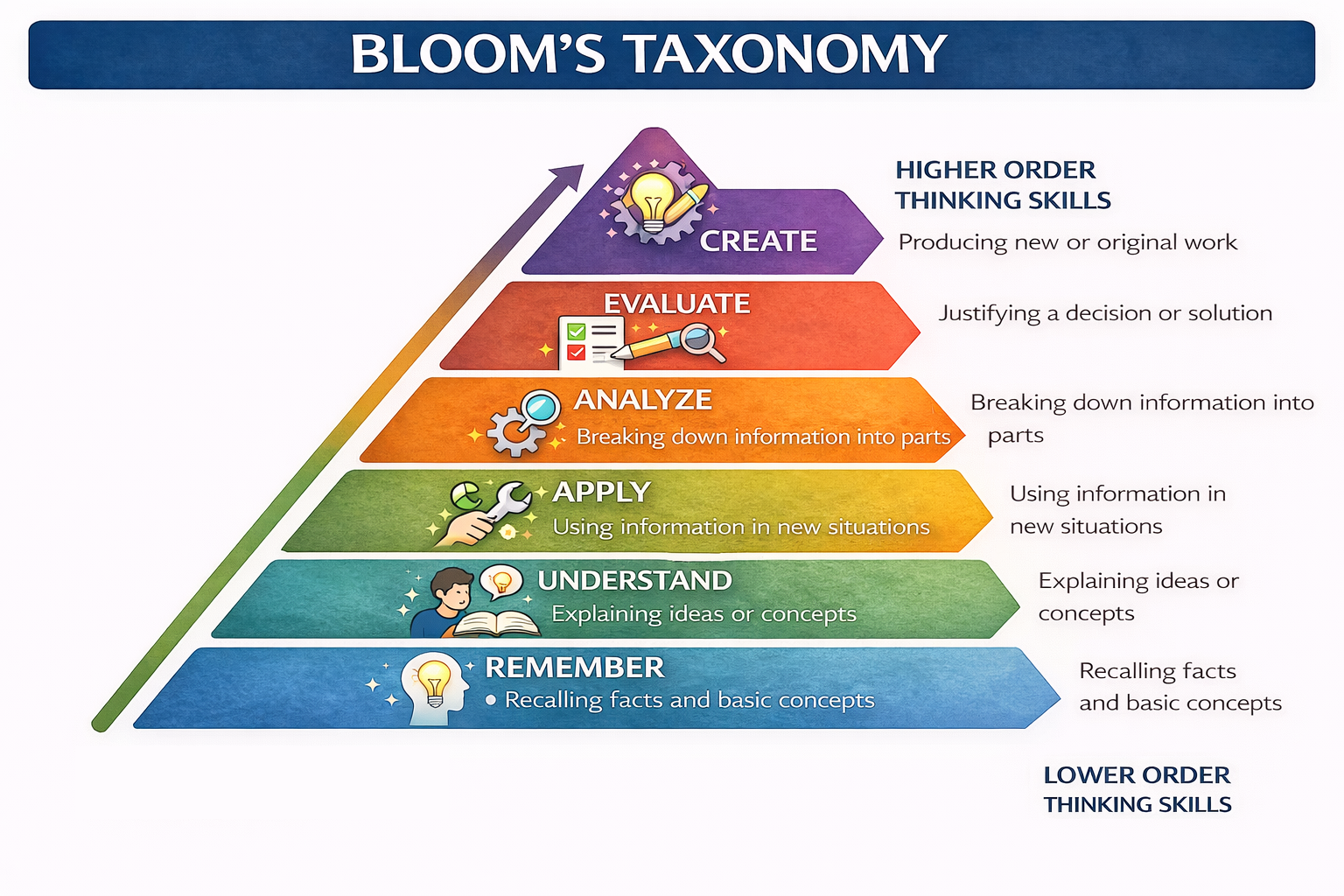}
\caption{Bloom's revised taxonomy~\citep{anderson2001revised}. Cognitive complexity increases from lower-order skills (remembering, understanding) to higher-order skills (evaluating, creating). Our module sequence follows this progression, with early modules at the base and the capstone module at the apex.}
\label{fig:blooms}
\end{figure}

Our design draws on four established frameworks:

\paragraph{Bloom's Taxonomy.} The revised taxonomy~\citep{bloom1956taxonomy,anderson2001revised} provides a progression from remembering through creating (Figure~\ref{fig:blooms}). Our 10-module sequence aligns with this progression: early modules emphasize understanding and applying Claude Code features, while later modules require analyzing, evaluating, and creating novel configurations.

\begin{figure}[t]
\centering
\includegraphics[width=0.85\textwidth]{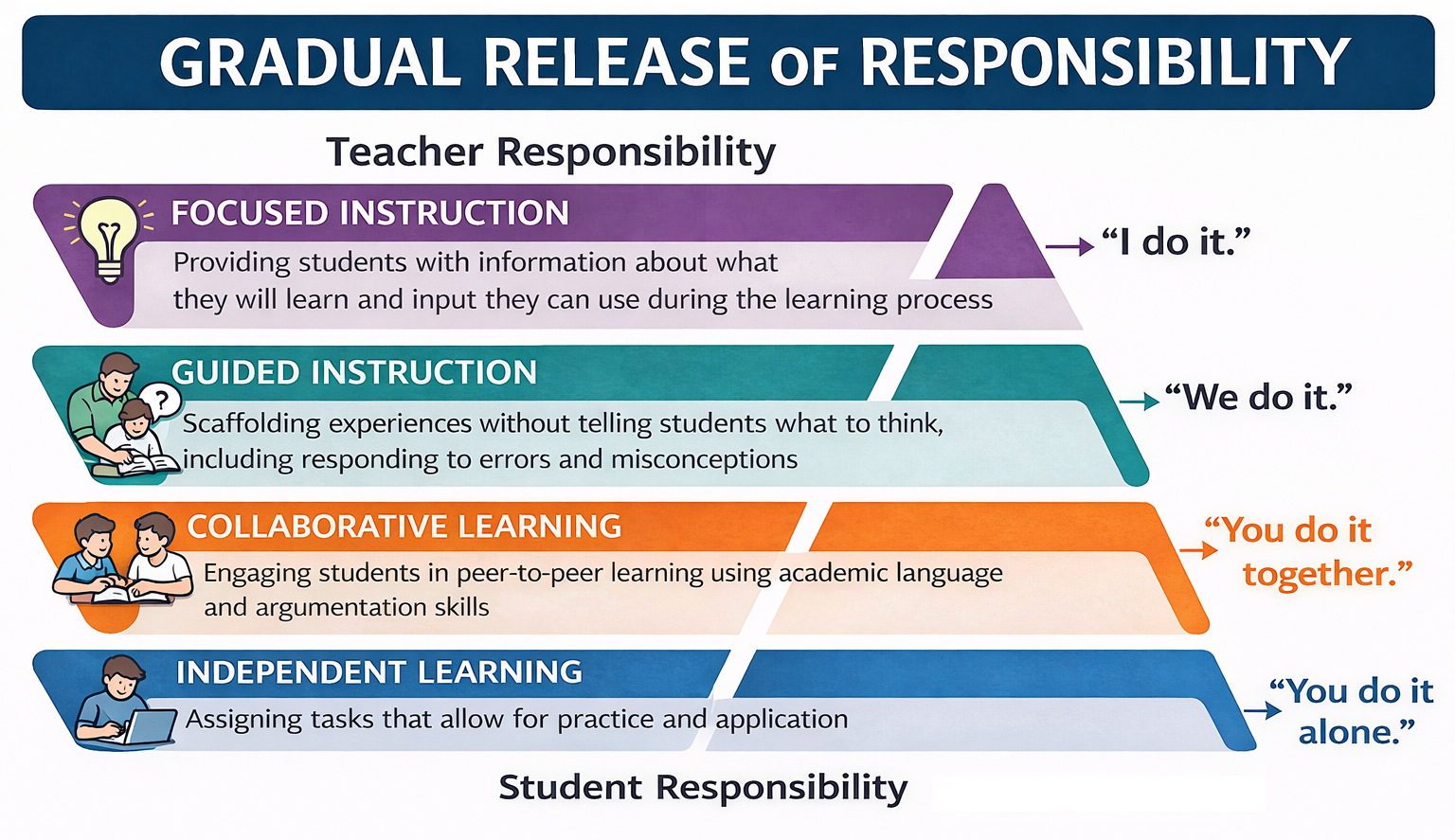}
\caption{The Gradual Release of Responsibility framework~\citep{fisher2013better}. Teacher responsibility decreases (top to bottom) as student responsibility increases across four phases: focused instruction, guided instruction, collaborative learning, and independent learning.}
\label{fig:grr}
\end{figure}

\paragraph{Gradual Release of Responsibility (GRR).} Fisher and Frey's framework~\citep{fisher2013better} describes a four-phase instructional model: focused instruction (``I do it''), guided instruction (``We do it''), collaborative learning (``You do it together''), and independent learning (``You do it alone''). Our persona progression (Guide $\to$ Collaborator $\to$ Peer $\to$ Launcher) is an adaptation of GRR for single-learner AI-mediated instruction, mapping 1:1 to its four phases. Guide corresponds to ``focused instruction'' (``I do it''): the AI provides information about what the learner will encounter and models how features work, though in our context the learner is always hands-on rather than passively observing. Collaborator maps to ``guided instruction'' (``We do it''), where the AI scaffolds without dictating. Peer maps to ``collaborative learning'' (``You do it together''), reinterpreted for a single-learner context: the AI adopts an equal-footing stance rather than facilitating peer group work. Launcher maps to ``independent learning'' (``You do it alone''). To our knowledge, this adaptation has not been previously applied to agentic tool instruction. This progression parallels the four-level agency model proposed by \citet{yan2025passive}, which describes AI roles escalating from adaptive instrument through proactive assistant and co-learner to peer collaborator, though their APCP framework addresses general human-AI collaboration rather than structured instruction. Figure~\ref{fig:grr} illustrates the framework and its mapping to our persona stages.

\paragraph{Cognitive Load Theory.} Sweller's framework~\citep{sweller1988cognitive} distinguishes intrinsic, extraneous, and germane cognitive load. Recent neuroscience evidence underscores this concern: \citet{kosmyna2025brain} found that unstructured AI assistance during writing tasks led to measurable cognitive debt, with users developing weaker neural engagement patterns over repeated sessions. Our step-pacing mechanism is designed to address the unique challenge that AI instructors, unlike human teachers, can generate unbounded volumes of instruction in a single response, risking extraneous load that may overwhelm working memory.

\paragraph{Constructionism.} Papert's constructionism~\citep{papert1980mindstorms} holds that learning is most effective when learners construct personally meaningful artifacts. Each of our five project options produces a functional software artifact (not a disposable exercise) that the learner retains and can extend after completing the curriculum.

\subsection{Auto-Updating Systems}

The concept of systems that monitor and improve their own outputs has gained traction in LLM research. Reflexion~\citep{shinn2023reflexion} introduces verbal reinforcement learning, where agents reflect on task feedback to improve subsequent attempts. Self-Refine~\citep{madaan2023selfrefine} demonstrates iterative self-improvement through self-generated feedback. Our sync pipeline (Section~\ref{sec:staying-current}) applies a related principle to educational content: the onboarding agent detects upstream releases, updates teaching steps, and verifies structural integrity before the learner begins instruction.

\section{System Architecture}
\label{sec:architecture}

\subsection{Repository Structure}

\begin{figure}[t]
\centering
\begin{tikzpicture}[
  every node/.style={font=\small},
  box/.style={draw, rounded corners=5pt, minimum width=2.6cm, minimum height=1.0cm, align=center, thick},
  pathbox/.style={draw, rounded corners=5pt, minimum width=1.6cm, minimum height=0.8cm, align=center, thick, fill=gray!8},
  modbox/.style={draw, rounded corners=3pt, minimum width=1.05cm, minimum height=0.55cm, font=\footnotesize\bfseries, align=center, thick, text=white},
  arrow/.style={-{Stealth[length=6pt]}, very thick, gray!70},
]

\coordinate (center) at (0,0);

\node[font=\bfseries\large] (title) at (center) {cc-self-train/};

\node[box, fill=blue!12] (dotclaude) at (-3.3,-1.2) {\textbf{.claude/}\\{\footnotesize config}};
\node[box, fill=yellow!20] (claudemd) at (0,-1.2) {\textbf{CLAUDE.md}\\{\footnotesize conventions}};
\node[box, fill=yellow!12] (readme) at (3.3,-1.2) {\textbf{README.md}\\{\footnotesize overview}};

\draw[arrow] (0,-1.85) -- (0,-2.55);

\node[box, fill=gray!12] (context) at (-1.65,-3.15) {\textbf{context/}\\{\footnotesize 22 docs}};
\node[box, fill=green!12] (tests) at (1.65,-3.15) {\textbf{tests/}\\{\footnotesize 8 suites}};

\draw[arrow] (0,-3.8) -- (0,-4.5);

\node[pathbox] (canvas) at (-3.6,-5.05) {Canvas};
\node[pathbox] (forge) at (-1.8,-5.05) {Forge};
\node[pathbox] (nexus) at (0,-5.05) {Nexus};
\node[pathbox] (sentinel) at (1.8,-5.05) {Sentinel};
\node[pathbox] (byop) at (3.6,-5.05) {BYOP};

\draw[arrow] (0,-5.6) -- (0,-6.2);

\node[font=\bfseries] (modlabel) at (0,-6.55) {10 Modules per Path};

\node[modbox, fill=guidecolor!85] (m1) at (-5.27,-7.2) {M1};
\node[modbox, fill=guidecolor!85] (m2) at (-4.1,-7.2) {M2};
\node[modbox, fill=guidecolor!85] (m3) at (-2.93,-7.2) {M3};
\node[modbox, fill=collabcolor!85] (m4) at (-1.76,-7.2) {M4};
\node[modbox, fill=collabcolor!85] (m5) at (-0.59,-7.2) {M5};
\node[modbox, fill=collabcolor!85] (m6) at (0.58,-7.2) {M6};
\node[modbox, fill=peercolor!85] (m7) at (1.75,-7.2) {M7};
\node[modbox, fill=peercolor!85] (m8) at (2.92,-7.2) {M8};
\node[modbox, fill=peercolor!85] (m9) at (4.09,-7.2) {M9};
\node[modbox, fill=launchercolor!85] (m10) at (5.26,-7.2) {M10};

\node[font=\footnotesize] (leg) at (0,-7.95) {%
  \textcolor{guidecolor}{\rule{0.7cm}{7pt}}~Guide \quad
  \textcolor{collabcolor}{\rule{0.7cm}{7pt}}~Collaborator \quad
  \textcolor{peercolor}{\rule{0.7cm}{7pt}}~Peer \quad
  \textcolor{launchercolor}{\rule{0.7cm}{7pt}}~Launcher%
};

\end{tikzpicture}
\caption{System architecture of \textsc{cc-self-train}. Five project paths share 10 modules with identical feature sequencing. Module colors indicate persona stage. The \texttt{.claude/} directory contains onboarding skills, session hooks, adaptive learning scripts, and the curriculum sync pipeline. The 22 context documents serve as reference material read by Claude Code during instruction.}
\label{fig:architecture}
\end{figure}
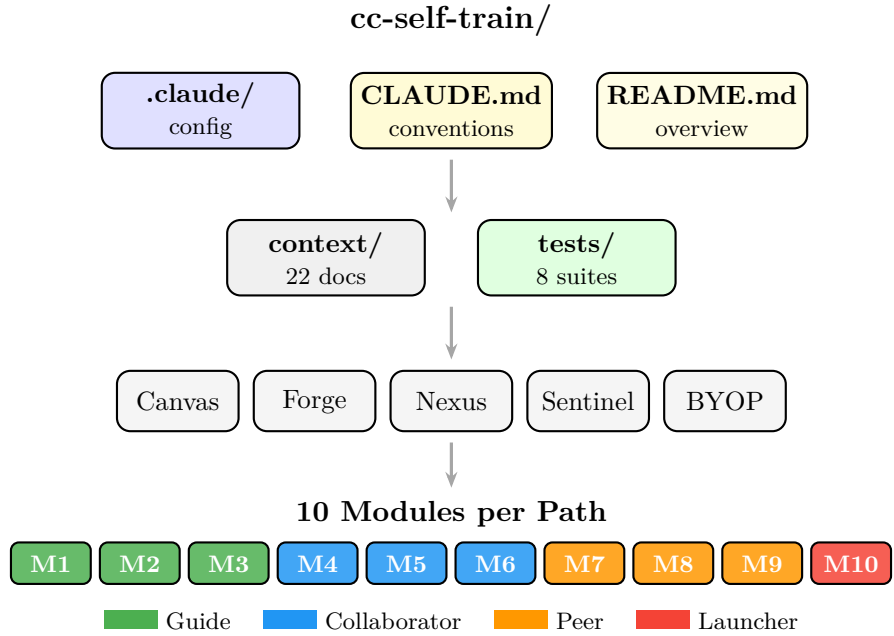

\textsc{cc-self-train} is organized as a single Git repository with the structure shown in Figure~\ref{fig:architecture}. The repository comprises:

\begin{itemize}[leftmargin=2em]
  \item \textbf{5 project directories} (\texttt{projects/canvas}, \texttt{forge}, \texttt{nexus}, \texttt{sentinel}, \texttt{byop}), each containing 10 module files and a project-specific README.
  \item \textbf{22 context documents} (\texttt{context/}) providing reference material for every Claude Code feature.
  \item \textbf{Configuration files} (\texttt{.claude/}) including the onboarding skill (\texttt{/start}), session hooks, adaptive learning observation and context-injection scripts, and sync pipeline.
  \item \textbf{8 test files} (\texttt{tests/}) with parametrized test suites.
  \item \textbf{A workspace directory} (\texttt{workspace/}, gitignored) where learner projects are scaffolded.
\end{itemize}

\subsection{The Five Learning Paths}

Each path produces a different software artifact while teaching the same 10 Claude Code feature sets in the same order:

\begin{enumerate}[leftmargin=2em]
  \item \textbf{Canvas} (Personal Portfolio Site): HTML/CSS/JS with no build tools. Recommended for first-time users due to zero toolchain friction.
  \item \textbf{Forge} (Personal Dev Toolkit): A CLI tool for notes, snippets, and templates. Language-agnostic.
  \item \textbf{Nexus} (Local API Gateway): A local server with routing, rate limiting, and caching. Language-agnostic.
  \item \textbf{Sentinel} (Code Analyzer \& Test Generator): A static analysis tool with auto-generated tests. Language-agnostic.
  \item \textbf{BYOP} (Bring Your Own Project): Learners apply the curriculum to an existing codebase, providing the ultimate transfer test.
\end{enumerate}

Canvas uses HTML/CSS/JS to eliminate toolchain setup. Forge, Nexus, and Sentinel span CLI tools, backend services, and developer tooling, respectively, allowing learners to choose based on interest rather than difficulty. BYOP, added in v2.12.0, tests whether learners can generalize Claude Code mastery to arbitrary codebases.

\subsection{Module Sequencing and Feature Dependencies}

The 10 modules follow a strict dependency order that builds compositional understanding (Table~\ref{tab:modules}). Each module assumes mastery of all prior modules. The sequencing reflects a mix of technical dependencies and conceptual prerequisites: guard rails (Module~7) extend the hook infrastructure introduced in Module~5; subagents (Module~8) reuse the Markdown-with-frontmatter conventions introduced with skills (Module~4); hooks (Module~5) are configured within the same \texttt{.claude/} directory structure introduced in Module~3; and the capstone module~(10) integrates all prior features.

\begin{table}[ht]
\centering
\caption{The 10-module curriculum with Claude Code features and teaching persona (intermediate schedule shown; beginners and advanced learners follow different persona boundaries, see Section~\ref{sec:persona}).}
\label{tab:modules}
\small
\begin{tabularx}{\textwidth}{c l X c}
\toprule
\textbf{\#} & \textbf{Module} & \textbf{CC Features (selected)} & \textbf{Persona} \\
\midrule
1 & Setup \& First Contact & \texttt{CLAUDE.md}, \texttt{/init}, \texttt{/memory}, interactive mode, keyboard shortcuts & Guide \\
2 & Blueprint \& Build & Plan mode, git integration, basic prompting, model selection & Guide \\
3 & Rules, Memory \& Context & \texttt{.claude/rules/}, \texttt{CLAUDE.local.md}, \texttt{@imports}, \texttt{/context}, \texttt{/compact} & Guide \\
4 & Skills \& Commands & \texttt{SKILL.md}, frontmatter, custom commands, hot-reload, argument substitution & Collaborator \\
5 & Hooks & \texttt{SessionStart}, \texttt{PostToolUse}, \texttt{Stop} hooks, matchers, hook scripting & Collaborator \\
6 & MCP Servers & MCP servers, \texttt{.mcp.json}, scopes, skills + MCP integration & Collaborator \\
7 & Guard Rails & \texttt{PreToolUse}, hook decision control, prompt-based hooks & Peer \\
8 & Subagents & \texttt{.claude/agents/}, subagent frontmatter, chaining, parallel, background & Peer \\
9 & Tasks \& TDD & Tasks system, dependencies, cross-session persistence, TDD loops & Peer \\
10 & Parallel Dev \& Eval & Worktrees, agent teams, plugins, evaluation framework, continuous learning & Launcher \\
\bottomrule
\end{tabularx}
\end{table}

\subsection{Onboarding as State Machine}

The entry point is the \texttt{/start} skill, a multi-step onboarding flow implemented as a state machine with resume capability. The flow proceeds through: curriculum version check $\to$ project selection $\to$ curriculum upgrade (if a newer release was detected) $\to$ OS detection $\to$ language selection (skipped for Canvas/BYOP) $\to$ experience level $\to$ progress resume (if prior state exists) $\to$ environment verification and tool installation $\to$ project scaffolding $\to$ Module~1 delivery. The resume check is deliberately placed after project and experience selection so that the system can validate prior state against the learner's current choices. The curriculum step operates in two phases: a quick version comparison before project selection, followed by the full sync pipeline (fetch changelog, triage changes, update module and context files) after project selection, since the sync needs to know which project path the learner chose. Each state is persisted to \texttt{.claude/onboarding-state.json}, enabling the onboarding to survive session interruptions, context compaction events, and even terminal crashes. This design decision arose from observed failure modes during early testing where learners would lose progress mid-onboarding.

\section{Persona Progression Model}
\label{sec:persona}

The persona progression model addresses a recurring mismatch in AI-mediated instruction: a static tone either underwhelms beginners or patronizes experts.

\subsection{Design Rationale}

Human instructors naturally shift their communication style as content becomes more advanced. Early lessons receive patient explanations; advanced topics receive terse pointers to relevant literature. This shift supports engagement across the learning trajectory. However, AI assistants typically operate with a fixed persona regardless of content stage. The persona progression model explicitly encodes this shift, tying instructional tone to module position in the sequence.

\subsection{The Four Stages}

The curriculum defines four personas, each specified as a prompt engineering directive in every module file (Figure~\ref{fig:persona}):

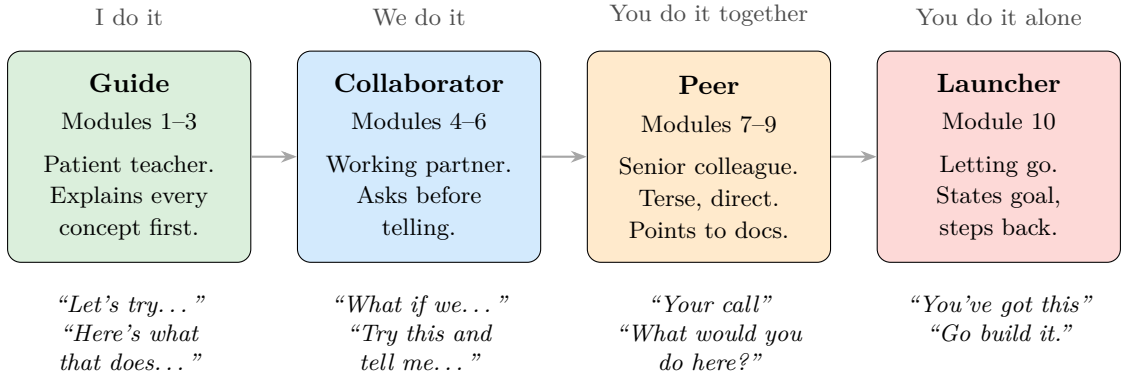
\begin{figure}[t]
\centering
\begin{tikzpicture}[
  node distance=0.3cm,
  stage/.style={draw, rounded corners=5pt, minimum width=3.2cm, minimum height=2.8cm, align=center, font=\small},
  arrow/.style={-{Stealth[length=6pt]}, thick, gray!70},
  phrase/.style={font=\footnotesize\itshape, text width=3.0cm, align=center},
]

\node[stage, fill=guidecolor!20] (guide) {%
  \textbf{Guide}\\[2pt]
  {\footnotesize Modules 1--3}\\[4pt]
  {\footnotesize Patient teacher.}\\
  {\footnotesize Explains every}\\
  {\footnotesize concept first.}%
};

\node[stage, fill=collabcolor!20, right=0.6cm of guide] (collab) {%
  \textbf{Collaborator}\\[2pt]
  {\footnotesize Modules 4--6}\\[4pt]
  {\footnotesize Working partner.}\\
  {\footnotesize Asks before}\\
  {\footnotesize telling.}%
};

\node[stage, fill=peercolor!20, right=0.6cm of collab] (peer) {%
  \textbf{Peer}\\[2pt]
  {\footnotesize Modules 7--9}\\[4pt]
  {\footnotesize Senior colleague.}\\
  {\footnotesize Terse, direct.}\\
  {\footnotesize Points to docs.}%
};

\node[stage, fill=launchercolor!20, right=0.6cm of peer] (launcher) {%
  \textbf{Launcher}\\[2pt]
  {\footnotesize Module 10}\\[4pt]
  {\footnotesize Letting go.}\\
  {\footnotesize States goal,}\\
  {\footnotesize steps back.}%
};

\draw[arrow] (guide) -- (collab);
\draw[arrow] (collab) -- (peer);
\draw[arrow] (peer) -- (launcher);

\node[phrase, below=0.3cm of guide] {``Let's try\ldots''\\``Here's what\\that does\ldots''};
\node[phrase, below=0.3cm of collab] {``What if we\ldots''\\``Try this and\\tell me\ldots''};
\node[phrase, below=0.3cm of peer] {``Your call''\\``What would you\\do here?''};
\node[phrase, below=0.3cm of launcher] {``You've got this''\\``Go build it.''};

\node[font=\footnotesize, text=gray!60!black, above=0.2cm of guide] {I do it};
\node[font=\footnotesize, text=gray!60!black, above=0.2cm of collab] {We do it};
\node[font=\footnotesize, text=gray!60!black, above=0.2cm of peer] {You do it together};
\node[font=\footnotesize, text=gray!60!black, above=0.2cm of launcher] {You do it alone};

\end{tikzpicture}
\caption{The four-stage persona progression with example phrases and mapping to the Gradual Release of Responsibility (GRR) framework~\citep{fisher2013better}. In our adaptation, ``I do it'' is reinterpreted: the AI models and explains while the learner executes, rather than the learner passively observing. Module ranges shown are for the intermediate schedule; beginners and advanced learners follow different boundaries (see Section~\ref{sec:persona}).}
\label{fig:persona}
\end{figure}

\paragraph{Stage 1: Guide (Modules 1--3).} The AI acts as a patient teacher. Every concept is explained before the learner is asked to use it. Technical terms are defined on first use. Small successes are acknowledged (``Nice, you just created your first rule file!''). When errors occur, the AI walks through the fix step by step. This stage corresponds to GRR's ``focused instruction'' phase (``I do it''): the AI provides information and models how features work, though the learner is always hands-on rather than passively observing. The learner is building foundational vocabulary and mental models for the tool's configuration system, git integration, and rule hierarchy.

\paragraph{Stage 2: Collaborator (Modules 4--6).} The AI shifts to a working partner. Basic concepts (git, CLAUDE.md, rules) are no longer re-explained. The AI asks questions before providing answers (``What do you think this hook should trigger on?''). Code examples become less complete, replaced by pointers (``The skill needs a frontmatter block. Check the docs if you need the format''). When errors occur, the AI asks ``What do you see in the error?'' before stepping in. This corresponds to GRR's ``guided instruction'' phase (``We do it''), where the AI scaffolds experiences without dictating what to think.

\paragraph{Stage 3: Peer (Modules 7--9).} The AI adopts a senior colleague's manner. Guidance is terse and direct. Rather than inline explanations, the AI points to reference files (``Check \texttt{context/hooks.txt} for the full event list''). Debugging is learner-driven; the AI intervenes only after the learner has attempted a fix. Challenges replace instructions: ``Can you wire this up without me spelling it out?'' This maps to GRR's ``collaborative learning'' phase (``You do it together''), reinterpreted for a single-learner context: the AI acts as an equal-footing partner rather than facilitating peer group work.

\paragraph{Stage 4: Launcher (Module 10).} The AI states the goal and steps back. Intervention occurs only after multiple failed attempts. Framing emphasizes the learner's existing competence: ``You already know how to do this.'' The final module ends with genuine recognition of mastery. This is GRR's ``independent learning'' phase.

\subsection{Implementation}

Each module file contains an explicit persona directive as a metadata line:

\begin{lstlisting}[language={},xleftmargin=2em]
**Persona -- Peer:** Terse guidance, point to docs,
let them debug first. "Your call", "What would you
do here?"
\end{lstlisting}

This line is consumed by Claude Code when reading the module file, shaping the tone and scaffolding depth of all subsequent responses within that module. A single line of natural language prompt engineering replaces what could otherwise be a more elaborate programmatic system. Persona adaptation works through descriptive prompt directives alone, without fine-tuning, RLHF, or RAG.

The persona boundaries are initially set at onboarding based on declared experience level, selecting from three schedules: beginners receive extended scaffolding (4/3/2/1 Guide/Collaborator/Peer/Launcher modules), intermediate learners follow the baseline progression (3/3/3/1), and advanced users reach peer-level autonomy earlier (1/3/5/1). However, these boundaries are not permanently fixed. The adaptive learning system (Section~\ref{sec:adaptive}) observes engagement quality during instruction and adjusts the learner's Effective Level at module boundaries, allowing the persona schedule to shift dynamically based on observed behavior. A second, orthogonal dimension controls explanation depth within whatever persona is active: instructions written to \texttt{CLAUDE.local.md} during onboarding determine how much background context accompanies each response. A beginner in Module 7 still receives the Peer tone but gets more thorough explanations of underlying concepts; an advanced user in Module 1 still receives the Guide tone but skips basic tool definitions. The persona governs \emph{how much hand-holding}; the experience level (and its runtime-adjusted Effective Level) governs both \emph{which persona is active} and \emph{how much background} is provided.

\subsection{Validation}

The persona specification is enforced by automated tests. The \texttt{TestAdaptivePersonaTable} class in \texttt{tests/test\_modules.py} verifies that experience-level persona boundaries cover all 10 modules without gaps and that personas never regress. A separate \texttt{TestPersonaProgression} class validates that each module file contains the correct persona directive for the intermediate schedule:

\begin{lstlisting}[language=Python,xleftmargin=2em]
ADAPTIVE_MAP = {
  "beginner":     {"Guide": range(1,5),
    "Collaborator": range(5,8),
    "Peer": range(8,10),  "Launcher": range(10,11)},
  "intermediate": {"Guide": range(1,4),
    "Collaborator": range(4,7),
    "Peer": range(7,10),  "Launcher": range(10,11)},
  "advanced":     {"Guide": range(1,2),
    "Collaborator": range(2,5),
    "Peer": range(5,10),  "Launcher": range(10,11)},
}
\end{lstlisting}

The test suite parametrically verifies that every experience level covers all 10 modules without gaps, that personas never regress, and that module-file defaults match the intermediate schedule. This ensures that curriculum updates, including those made by the automated sync pipeline, cannot accidentally violate the persona progression for any experience level.

\section{Adaptive Learning System}
\label{sec:adaptive}

The persona schedules described in Section~\ref{sec:persona} adapt scaffolding across the curriculum, but the schedule is selected once at onboarding based on self-reported experience. Self-reports can be inaccurate: a developer who has used GitHub Copilot extensively may overestimate their readiness for an agentic tool, while a newcomer with strong software engineering fundamentals may underestimate theirs. A recent randomized controlled trial with 770 high school students found that adaptive problem sequencing improved exam performance by 0.15 standard deviations ($p < 0.05$), equivalent to 6--9 months of additional schooling~\citep{chung2025adaptive}. The key finding: gains were almost entirely mediated by increased \emph{engagement quality}, not harder problems or more problems. Beginners benefited most (0.215 SD); experienced students saw negligible effect. This motivated a lightweight observation layer that tracks engagement quality and feeds back into persona selection.

\subsection{Three-Layer Architecture}

The adaptive learning system operates through three layers, each implemented as a Claude Code hook or instruction set.

\paragraph{Layer 1: Observation.} A Stop hook (\texttt{observe-interaction.js}) fires silently after every Claude response. It reads the student's most recent message from the conversation transcript and classifies it into one of six engagement categories using keyword heuristics (Table~\ref{tab:engagement}). Each category carries a quality score from 1 to 5. Scores accumulate in a local file (\mbox{\texttt{learner-profile.json}}, gitignored) that tracks both lifetime and per-module interaction counts, a rolling quality average, and a trend indicator (improving, stable, or declining) computed by comparing recent scores against preceding ones. In addition to aggregate statistics, the script maintains a sliding window of the last five non-neutral categories and computes two streak booleans: \texttt{struggleStreak} (three or more consecutive \texttt{answer\_seeking} or \texttt{passive\_acceptance} classifications) and \texttt{engagementStreak} (three or more consecutive \texttt{concept\_question} or \texttt{independent\_exploration} classifications). This streak detection is motivated by research showing that non-mastery rules (repeated incorrect responses) carry higher gradient sensitivity than mastery rules in neural-symbolic knowledge tracing models, meaning sequential failure patterns are more informative than isolated events~\citep{hooshyar2026neural}. The script is silent (no output to the student), adds negligible latency, and uses a filesystem lock to prevent re-entrancy.

\begin{table}[h]
\centering
\caption{Engagement categories used by the observation hook. Quality scores range from 1 (unproductive) to 5 (highly productive).}
\label{tab:engagement}
\small
\begin{tabular}{llcl}
\toprule
\textbf{Category} & \textbf{Type} & \textbf{Score} & \textbf{Example Signals} \\
\midrule
\texttt{concept\_question}        & Productive   & 5 & ``why does\ldots'', ``how does\ldots'', ``explain'' \\
\texttt{independent\_exploration} & Productive   & 4 & ``I tried\ldots'', ``I noticed\ldots'', ``I figured out'' \\
\texttt{debug\_attempt}           & Productive   & 3 & ``error'', ``not working'', describing observed behavior \\
\texttt{neutral}                  & ---          & 3 & Navigation, ``next module'', unmatched input \\
\texttt{answer\_seeking}          & Unproductive & 1 & ``just do it'', ``write it for me'', ``give me the code'' \\
\texttt{passive\_acceptance}      & Unproductive & 1 & Very short reply ($<$15 chars) to a long response \\
\bottomrule
\end{tabular}
\end{table}

\paragraph{Layer 2: Context Injection.} A SessionStart hook (\texttt{learner-context.js}) reads the learner profile at the beginning of each session. If the profile contains at least five non-neutral interactions, the hook computes the productive/unproductive ratio and injects a short teaching note into Claude's context. The note is invisible to the student and includes the current engagement trend, dominant interaction pattern, and a concrete teaching directive. Three tiers govern the directive: a productive ratio $\geq 0.7$ encourages curiosity and deeper exploration; a ratio between 0.4 and 0.7 redirects answer-seeking behavior with guiding questions; a ratio below 0.4 provides additional structure and scaffolding. When a streak is active, the hook appends a priority alert: a struggle streak produces \emph{``Offer more scaffolding NOW''} while an engagement streak produces \emph{``Student is in flow. Match with deeper content.''}\ Only one streak type is surfaced per session to prevent conflicting signals. This gives Claude session-to-session memory of how the student learns without the student needing to say anything.

\paragraph{Layer 3: Adaptation.} At module boundaries (when the student requests the next module), instructions in \texttt{CLAUDE.md} direct the AI to read the accumulated engagement signals and decide whether to adjust the student's \emph{Effective Level}, a new field in \texttt{CLAUDE.local.md} that can diverge from the self-reported experience level:

\begin{itemize}[leftmargin=2em]
  \item If \texttt{moduleAverageQuality} $\geq 3.8$ \textbf{and} productive interactions (concept questions, independent exploration, and debug attempts) exceed 60\% of non-neutral interactions: bump the Effective Level \emph{up} one notch (e.g., beginner $\to$ intermediate for persona lookup).
  \item If \texttt{moduleAverageQuality} $\leq 2.0$ \textbf{and} unproductive interactions (answer-seeking + passive acceptance) exceed 50\%: bump \emph{down} one notch.
  \item Otherwise: hold steady.
\end{itemize}

\noindent The Effective Level feeds into the existing persona progression table (Section~\ref{sec:persona}), so a self-reported beginner who consistently asks deep conceptual questions may receive Collaborator-level teaching earlier than the default schedule. Conversely, a self-reported intermediate who is struggling receives additional Guide-level scaffolding. Changes are silent (the student is never told their level shifted), limited to one notch per module boundary, and module-level counters reset after each transition.

\paragraph{Two-Timescale Adaptation.} The streak booleans from Layer~1 and the Effective Level from Layer~3 create a two-timescale system. The \emph{slow clock} (Effective Level) adjusts the persona schedule at module boundaries using aggregate statistics. The \emph{fast clock} (streak detection) triggers scaffolding changes mid-module using sequential patterns. An \emph{asymmetric response principle} governs the fast clock: the system is quicker to increase scaffolding (responding to struggle streaks immediately) than to withdraw it. This reflects a pedagogical heuristic grounded in the Hooshyar et al.\ finding: a student who is breezing through may simply be on an easy topic, but three consecutive answer-seeking messages is a strong signal of genuine difficulty. Over-scaffolding a strong student briefly costs little; under-scaffolding a struggling student risks permanent disengagement.

\subsection{Design Constraints}

Three deliberate constraints shape the system. First, the observation layer uses keyword heuristics rather than LLM-based classification to avoid per-turn inference costs and latency; individual classifications are noisy, but aggregated over dozens of interactions per module the signal is clear. Second, \emph{Effective Level} changes occur only at module boundaries to avoid disorienting the learner with intra-module persona shifts; streak-based scaffolding adjustments are the sole exception, and these modulate explanation depth rather than switching the active persona. Third, level changes are silent: informing the student of a demotion could be discouraging, and informing them of a promotion could create self-consciousness about maintaining it.

\subsection{Relation to Prior Work}

The Chung et al.\ finding that engagement quality, not problem difficulty, mediates learning gains directly shaped the scoring function: categories reflecting active reasoning (concept questions, independent exploration) receive high scores, while categories reflecting passive consumption receive low scores. The Hooshyar et al.\ finding that non-mastery rules carry higher gradient sensitivity than mastery rules in neural-symbolic knowledge tracing models~\citep{hooshyar2026neural} motivated both the streak detection mechanism and the asymmetric response principle: the system treats consecutive struggle signals as a stronger indicator than consecutive engagement signals, responding to the former immediately rather than waiting for aggregate statistics to shift. The architecture is deliberately lightweight compared to full Intelligent Tutoring Systems that maintain complex student models via knowledge tracing or POMDPs with particle filtering. The system trades model sophistication for deployability: it runs entirely through Claude Code's hook system with no external dependencies, no API calls for observation, and no web backend.

\section{Cross-Domain Transfer Design}
\label{sec:crossdomain}

\subsection{The Unified Curriculum Constraint}

A core design decision: all five paths teach the same 10 Claude Code feature sets in the same order, designed to facilitate cross-domain transfer of tool mastery.

The constraint means that a learner who completes the Canvas path and then begins Forge should encounter familiar feature progression in an unfamiliar domain context. The Claude Code skills are intended to transfer; only the project-specific application changes. This mirrors how professional developers use tools: the same Git workflow applies whether building a web app, a CLI tool, or a microservice.

\subsection{Domain Adaptation}

While the feature sequence is fixed, each module is adapted to its project domain. Consider Module~7 (Guard Rails), which teaches PreToolUse hooks with decision control. The core Claude Code features are identical across all paths (\texttt{permissionDecision}, \texttt{additionalContext}, and \texttt{updatedInput}), but the exercises differ:

\begin{itemize}[leftmargin=2em]
  \item \textbf{Canvas:} A hook that blocks HTML writes missing \texttt{alt} attributes on images (accessibility enforcement).
  \item \textbf{Forge:} A hook that validates storage JSON file structure before writes.
  \item \textbf{Nexus:} A hook that checks API route configurations for required fields and security risks.
  \item \textbf{Sentinel:} A hook that validates analysis rule definitions for correctness.
  \item \textbf{BYOP:} A hook targeting a quality concern in the learner's own codebase.
\end{itemize}

Each exercise teaches the same three hook mechanisms (deny, inject context, and modify input) but contextualizes them within the project's domain, aiming to reinforce both Claude Code mastery and domain-relevant best practices.

\subsection{Feature Coverage Matrix}

The feature coverage matrix (Table~\ref{tab:features}) documents that all major Claude Code features currently covered are taught in all 5 paths. This matrix is enforced by the \texttt{TestModuleConsistencyAcrossProjects} test class, which extracts the \texttt{**CC features:**} line from every module file and verifies symmetric coverage across all projects. A feature added to one path's module that is not added to the other four will fail the test suite.

\begin{table}[ht]
\centering
\caption{Feature coverage matrix (abridged). All currently covered CC features are taught in all 5 paths. Full matrix available in the repository README. \checkmark{} indicates the feature is covered in that path's module.}
\label{tab:features}
\small
\begin{tabular}{l c c c c c c}
\toprule
\textbf{Feature} & \textbf{Canvas} & \textbf{Forge} & \textbf{Nexus} & \textbf{Sentinel} & \textbf{BYOP} & \textbf{Mod.} \\
\midrule
CLAUDE.md, /init, /memory & \checkmark & \checkmark & \checkmark & \checkmark & \checkmark & 1 \\
Plan mode, git integration & \checkmark & \checkmark & \checkmark & \checkmark & \checkmark & 2 \\
.claude/rules/, @imports & \checkmark & \checkmark & \checkmark & \checkmark & \checkmark & 3 \\
Skills (SKILL.md, frontmatter) & \checkmark & \checkmark & \checkmark & \checkmark & \checkmark & 4 \\
Hooks (SessionStart, PostToolUse) & \checkmark & \checkmark & \checkmark & \checkmark & \checkmark & 5 \\
MCP servers, .mcp.json & \checkmark & \checkmark & \checkmark & \checkmark & \checkmark & 6 \\
PreToolUse, decision control & \checkmark & \checkmark & \checkmark & \checkmark & \checkmark & 7 \\
Subagents, chaining, parallel & \checkmark & \checkmark & \checkmark & \checkmark & \checkmark & 8 \\
Tasks, TDD, dependencies & \checkmark & \checkmark & \checkmark & \checkmark & \checkmark & 9 \\
Worktrees, plugins, eval & \checkmark & \checkmark & \checkmark & \checkmark & \checkmark & 10 \\
\bottomrule
\end{tabular}
\end{table}

\subsection{BYOP as Transfer Test}

The Bring Your Own Project (BYOP) path, added in v2.12.0, serves as the ultimate transfer test. Unlike the four tutorial paths where exercises are pre-designed, BYOP requires learners to identify appropriate applications of each Claude Code feature within their own codebase. For example, Module~7's guard rail exercises require the learner to identify a quality concern in their project that warrants a PreToolUse hook; there is no pre-specified exercise. This demands genuine transfer of abstract tool knowledge to a concrete, unfamiliar context.

\section{Step-Pacing and Information Load Management}
\label{sec:pacing}

\subsection{The Information Overload Problem}

AI-as-instructor introduces a cognitive load challenge that has no direct parallel in human instruction. A human teacher naturally paces delivery through conversational turn-taking, reading body language, and responding to student confusion. An AI agent, when instructed to ``teach Module 5,'' can generate thousands of words of instruction in a single response, covering every step, explanation, and exercise without pause. This produces extraneous cognitive load~\citep{sweller1988cognitive} that overwhelms working memory and inhibits learning.

The problem is compounded by the terminal environment in which Claude Code operates. Unlike a graphical IDE with tabs and panels, a terminal session presents information as a linear text stream. A 3,000-word response scrolls past the visible area, requiring the learner to scroll back and forth to follow multi-step instructions while simultaneously executing commands.

\subsection{STOP Blocks as Pacing Primitives}

We introduce \emph{STOP blocks}, explicit pause directives embedded in module files that create hard boundaries in instruction delivery:

\begin{lstlisting}[language={},xleftmargin=2em]
**STOP -- What you just did:** You created a guard that
*prevents* Claude from writing inaccessible HTML.
Unlike the PostToolUse validator from Module 5 that
reports issues after the file is already written, this
PreToolUse hook blocks the write entirely.
\end{lstlisting}

A STOP block serves three functions: (1)~it forces the AI to pause and wait for the learner's response before continuing; (2)~it provides a reflection prompt intended to consolidate learning from the preceding step; and (3)~it creates a natural checkpoint where the learner can assess their understanding before moving forward.

The pacing rule is enforced through a directive in \texttt{CLAUDE.md} that instructs the AI agent to halt after each STOP block. Because enforcement relies on the LLM's adherence to prompt instructions rather than a hard system lock, occasional violations are possible, though in practice we have found compliance to be reliable:

\begin{quote}
\itshape Deliver one step at a time. Each numbered step is a separate message. After completing a step, STOP and wait for the user to respond before continuing to the next step. If a step has a STOP block, that is a hard boundary. Do not continue past it under any circumstances.
\end{quote}

\subsection{Cross-Session State Tracking}

Step progress is tracked in \texttt{CLAUDE.local.md}, a per-session state file that persists across context compaction events and session restarts. After each step, the AI updates the current step marker:

\begin{lstlisting}[language={},xleftmargin=2em]
Current Module: 7
Current Step: 7.3
\end{lstlisting}

This tracking mechanism addresses a unique challenge in LLM-based instruction: context window limits. When Claude Code's context approaches capacity, it automatically compresses prior messages. Without external state tracking, the AI would lose its place in the curriculum. The \texttt{CLAUDE.local.md} file survives compaction because it is re-read at the start of each interaction, serving as a persistent breadcrumb.

The system also includes a pre-advancement check: before proceeding to the next module, the AI must cross-reference the current step against the module's Checkpoint section. If any steps were skipped (due to debugging tangents, user digressions, or context compaction), the AI returns to cover the missing steps before advancing. The learner does not need to track their own progress; the checkpoint structure does.

\section{Automated Quality Assurance}
\label{sec:qa}

In the absence of runtime telemetry, automated testing provides our primary structural quality assurance. The test suite validates structural consistency across all 50 module files, using this as a proxy for pedagogical invariants.

\subsection{Test Suite Overview}

Table~\ref{tab:tests} summarizes the 8 test files and their coverage domains.

\begin{table}[ht]
\centering
\caption{Automated test suite summary. Tests are parametrized across 5 projects and 10 modules, yielding coverage across all 50 module files. Test counts reflect unique test functions before parametrization.}
\label{tab:tests}
\small
\begin{tabularx}{\textwidth}{l X c}
\toprule
\textbf{Test File} & \textbf{What It Validates} & \textbf{Functions} \\
\midrule
\texttt{test\_modules.py} & Module completeness, H1 titles, persona lines, CC features, numbering, persona progression, adaptive persona table, cross-project feature consistency & 15 \\
\texttt{test\_file\_structure.py} & Repository structure, required directories, required files, no unexpected files & -- \\
\texttt{test\_context.py} & Context file existence, non-empty content, expected file list & -- \\
\texttt{test\_hooks.py} & Hook configuration, script existence, expected hook events & -- \\
\texttt{test\_skill.py} & Onboarding skill structure, frontmatter, step completeness & -- \\
\texttt{test\_other\_skills.py} & Non-onboarding skills (sync, release) structure and metadata & -- \\
\texttt{test\_cross\_refs.py} & Cross-references between modules, context files, and README & -- \\
\texttt{conftest.py} & Shared fixtures, project list, module file list, constants & -- \\
\bottomrule
\end{tabularx}
\end{table}

\subsection{Parametrized Testing as Coverage Multiplier}

The key testing strategy is parametrization across the two-dimensional space of projects and modules. A single test function like \texttt{test\_persona\_matches\_module} is parametrized over 5 projects $\times$ 10 modules = 50 test cases. This approach ensures that every structural invariant is verified for every module file in the repository, making it infeasible for a curriculum update to violate consistency without triggering a test failure.

The parametrization is defined in \texttt{conftest.py}:

\begin{lstlisting}[language=Python,xleftmargin=2em]
PROJECTS = ["canvas", "forge", "nexus", "sentinel", "byop"]
MODULE_FILES = [
    "01-setup.md", "02-blueprint.md",
    "03-rules-memory-context.md", "04-skills-commands.md",
    "05-hooks.md", "06-mcp-servers.md",
    "07-guard-rails.md", "08-subagents.md",
    "09-tasks-tdd.md", "10-parallel-plugins-eval.md",
]
\end{lstlisting}

\subsection{Tests as Quality Proxy}

Beyond the pilot evaluation (Section~\ref{sec:evaluation}), our test suite proxies for ongoing quality by enforcing invariants that are necessary (though not sufficient) conditions for a well-structured curriculum:

\begin{itemize}[leftmargin=2em]
  \item \textbf{Completeness:} Every project has all 10 modules; no extra files exist.
  \item \textbf{Structural consistency:} Every module starts with a properly numbered heading, lists CC features, and specifies a persona.
  \item \textbf{Pedagogical consistency:} Persona assignments follow the prescribed progression across all projects.
  \item \textbf{Cross-project parity:} The CC features taught in each module are identical across all 5 projects.
\end{itemize}

These tests do not evaluate whether the instruction is effective (that would require user data). They enforce structural properties that we consider prerequisites for effective instruction, particularly important given that the curriculum auto-updates when new tool versions are detected (Section~\ref{sec:staying-current}).

\section{Staying Current}
\label{sec:staying-current}

AI coding tools ship breaking changes on a cadence measured in days. A curriculum that teaches last week's features is teaching the wrong things. Rather than treating content decay as a maintenance burden, \textsc{cc-self-train} is designed to stay current automatically: when a learner begins the curriculum, the onboarding agent checks whether its teaching materials cover the learner's installed version of Claude Code. If not, it updates the modules before instruction begins, so the learner always receives a current snapshot. The same pipeline can also be invoked standalone by a maintainer for bulk updates across all five project variants.

\subsection{The Sync Pipeline}

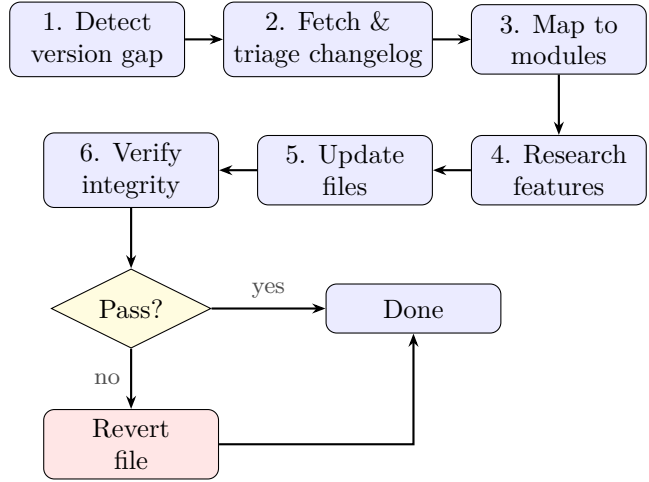
\begin{figure}[t]
\centering
\begin{tikzpicture}[
  node distance=0.5cm and 0.5cm,
  pipestep/.style={draw, rounded corners=4pt, fill=blue!8, minimum width=2.3cm, minimum height=0.65cm, align=center, font=\small},
  decision/.style={draw, diamond, aspect=2, fill=yellow!15, minimum width=1.5cm, align=center, font=\small},
  arrow/.style={-{Stealth[length=5pt]}, thick},
  lbl/.style={font=\footnotesize, text=gray!60!black},
]

\node[pipestep] (detect) {1. Detect\\version gap};
\node[pipestep, right=0.5cm of detect] (fetch) {2. Fetch \&\\triage changelog};
\node[pipestep, right=0.5cm of fetch] (map) {3. Map to\\modules};

\node[pipestep, below=0.8cm of map] (research) {4. Research\\features};
\node[pipestep, left=0.5cm of research] (update) {5. Update\\files};
\node[pipestep, left=0.5cm of update] (verify) {6. Verify\\integrity};

\node[decision, below=0.8cm of verify] (pass) {Pass?};
\node[pipestep, right=1.5cm of pass] (done) {Done};
\node[pipestep, below=0.8cm of pass, fill=red!10] (revert) {Revert\\file};

\draw[arrow] (detect) -- (fetch);
\draw[arrow] (fetch) -- (map);
\draw[arrow] (map) -- (research);
\draw[arrow] (research) -- (update);
\draw[arrow] (update) -- (verify);
\draw[arrow] (verify) -- (pass);
\draw[arrow] (pass) -- node[above, lbl] {yes} (done);
\draw[arrow] (pass) -- node[left, lbl] {no} (revert);
\draw[arrow] (revert.east) -- (revert.east -| done.south) -- (done.south);

\end{tikzpicture}
\caption{The sync pipeline (\texttt{/sync} skill). The pipeline detects upstream Claude Code releases, triages changelog entries, appends new steps to affected modules, and verifies structural integrity. Files that fail verification are reverted. During onboarding, only the learner's chosen project is updated; standalone invocation updates all five variants.}
\label{fig:sync}
\end{figure}

The update logic is implemented as a Claude Code skill, \texttt{/sync} (Figure~\ref{fig:sync}). Both invocation modes share the same six-step pipeline. When the tool has released new features since the curriculum was last updated, the pipeline:

\begin{enumerate}[leftmargin=2em]
  \item \textbf{Detects} the version gap between the curriculum's last-synced version and the latest Claude Code release.
  \item \textbf{Fetches} the upstream changelog and triages entries (keeping Added/Changed/Removed, skipping bug fixes).
  \item \textbf{Maps} relevant features to affected modules and context files using a predefined feature-to-module mapping.
  \item \textbf{Researches} new features via web search, official documentation, or changelog text.
  \item \textbf{Updates} affected module and context files, appending new steps before the Checkpoint section and matching the teaching persona depth for each target module.
  \item \textbf{Verifies} all modified files for structural integrity (step numbering, checkpoint existence, STOP block preservation), reverting any file that fails.
\end{enumerate}

The two modes differ in trigger and scope:

\begin{itemize}[leftmargin=2em]
  \item \textbf{Onboarding (automatic):} When a learner runs \texttt{/start}, the onboarding agent executes the pipeline if it detects a version gap. Only the learner's chosen project variant is updated, so the sync completes quickly and instruction can begin with current materials.
  \item \textbf{Standalone (manual):} A maintainer invokes \texttt{/sync} directly to update all five project variants in bulk, then reviews and commits the results.
\end{itemize}

The safe-append rule: for module files, the pipeline \emph{never modifies or renumbers existing steps}. New content is inserted immediately before the Checkpoint section at the end of each module. This ensures that learners who are mid-curriculum will not encounter renumbered or altered steps after a sync, and that the step-tracking system (Section~\ref{sec:pacing}) remains valid across updates. Context files (reference documentation in \texttt{context/}) are updated in-place, since learners do not step through them.

\subsection{Implications}

The auto-updating design has two implications:

\begin{enumerate}[leftmargin=2em]
  \item \textbf{Curriculum currency:} The AI coding tool landscape evolves on timescales of days or weeks rather than months or years, making static curricula obsolete quickly. Because the onboarding agent checks for upstream changes at the start of each learner's journey, new features are incorporated into the curriculum at the point of use. The learner always receives teaching materials that match their installed tool version.

  \item \textbf{Scalability:} The sync pipeline scales to any number of project variants without proportional human effort. Adding BYOP as a 5th path required generating 10 module files, which the pipeline then maintains alongside the original 40.
\end{enumerate}

\section{Pilot Evaluation}
\label{sec:evaluation}

To assess whether the curriculum produces measurable self-efficacy gains, we conducted a within-subjects pilot study with pre- and post-training self-assessments.

\subsection{Methodology}

\paragraph{Participants.} We recruited 27 professional software engineers from a single client organization. Participation was voluntary, and participants knew the author designed the curriculum. Participants self-reported their prior experience with Claude Code at one of three levels: beginner ($n=9$), intermediate ($n=8$), or advanced ($n=10$). All participants completed the full curriculum through the onboarding agent and at least one project path.

\paragraph{Instrument.} We designed a 10-item self-efficacy questionnaire mapped to the curriculum's feature progression (full item text in Appendix~\ref{app:survey}). Each item used a 5-point Likert scale (1 = Strongly Disagree, 5 = Strongly Agree). The items assessed two categories of skill:

\begin{itemize}[leftmargin=2em]
  \item \textbf{Foundational skills} (taught in early modules): explaining Claude Code's purpose, completing end-to-end changes without step-by-step guidance, understanding and using project memory (CLAUDE.md), and knowing when to plan versus execute immediately.
  \item \textbf{Advanced skills} (taught in later modules): understanding and creating custom skills (slash commands), understanding and setting up hooks for automated workflows, and troubleshooting off-track agent behavior by adjusting context, instructions, or memory.
\end{itemize}

\paragraph{Procedure.} Participants completed the pre-survey before beginning any curriculum module and the post-survey after completing their final module. The same 10 items appeared in both surveys. Surveys were self-administered via Google Forms with anonymized user IDs to enable paired analysis.

\paragraph{Analysis.} We used paired $t$-tests (two-tailed) for each item and overall, with Cohen's $d$ as the effect size measure. We also performed a Wilcoxon signed-rank test as a non-parametric confirmation of the overall result.

\subsection{Results}

Table~\ref{tab:evaluation} presents per-item results. All 10 items showed statistically significant gains ($p < 0.001$), with large effect sizes across the board (Cohen's $d$ range: 0.89--2.79).

\begin{table}[ht]
\centering
\caption{Pre/post self-efficacy ratings ($n = 27$). All gains are statistically significant at $p < 0.001$ (paired $t$-test, two-tailed). Scale: 1 = Strongly Disagree, 5 = Strongly Agree.}
\label{tab:evaluation}
\small
\begin{tabularx}{\textwidth}{X c c c c c}
\toprule
\textbf{Skill Area} & \textbf{Pre} & \textbf{Post} & \textbf{Gain} & \textbf{$t$(26)} & \textbf{$d$} \\
\midrule
\multicolumn{6}{l}{\textit{Foundational skills}} \\
Explain what CC is and when useful & 3.63 & 4.37 & +0.74 & 7.32 & 1.41 \\
End-to-end change without help & 3.44 & 4.33 & +0.89 & 5.77 & 1.11 \\
Understand project memory & 3.07 & 3.59 & +0.52 & 4.65 & 0.89 \\
Use memory across sessions & 2.81 & 3.81 & +1.00 & 7.65 & 1.47 \\
Plan vs.\ execute immediately & 3.52 & 4.04 & +0.52 & 5.29 & 1.02 \\
\midrule
\multicolumn{6}{l}{\textit{Advanced skills}} \\
Understand skills (slash commands) & 3.15 & 3.81 & +0.67 & 6.24 & 1.20 \\
Create and use a skill & 2.63 & 3.63 & +1.00 & 13.25 & 2.55 \\
Understand hooks & 2.41 & 3.78 & +1.37 & 11.32 & 2.18 \\
Set up a hook & 1.96 & 3.04 & +1.07 & 14.50 & 2.79 \\
Troubleshoot off-track behavior & 3.04 & 3.67 & +0.63 & 5.79 & 1.11 \\
\midrule
\textbf{Overall} & \textbf{2.97} & \textbf{3.81} & \textbf{+0.84} & \textbf{22.20} & \textbf{1.35} \\
\bottomrule
\end{tabularx}
\end{table}

The overall gain was confirmed by a Wilcoxon signed-rank test ($W = 0.0$, $p < 0.00001$), indicating that 26 of 27 participants improved (one advanced user scored at ceiling on both administrations).

\paragraph{Gains by expertise level.} Table~\ref{tab:gains-expertise} breaks down gains by self-reported experience. Beginners showed the largest gains (+1.18), intermediates were close behind (+0.94), and advanced users showed smaller but still significant gains (+0.46). This pattern is consistent with ceiling effects: advanced users started closer to the scale maximum, leaving less room for measured improvement. All three subgroups reached $p < 0.001$.

\begin{table}[ht]
\centering
\caption{Mean self-efficacy gains by expertise level. All subgroups show significant improvement ($p < 0.001$).}
\label{tab:gains-expertise}
\small
\begin{tabular}{l c c c c c}
\toprule
\textbf{Level} & \textbf{$n$} & \textbf{Pre} & \textbf{Post} & \textbf{Gain} & \textbf{$t$} \\
\midrule
Beginner & 9 & 1.61 & 2.79 & +1.18 & 22.60 \\
Intermediate & 8 & 2.91 & 3.85 & +0.94 & 14.36 \\
Advanced & 10 & 4.23 & 4.69 & +0.46 & 6.15 \\
\bottomrule
\end{tabular}
\end{table}

\subsection{Interpretation}

Two patterns stand out. First, the largest effect sizes appear on \emph{advanced features}: setting up hooks ($d = 2.79$), creating skills ($d = 2.55$), and understanding hooks ($d = 2.18$). These are the features taught in later modules where the persona has shifted from Guide to Collaborator or Peer, suggesting that the progressive scaffolding may be especially valuable for compositional features that are difficult to learn from documentation alone.

Second, the expertise-level gradient supports the curriculum's experience-stratified persona boundaries. Beginners, who receive more Guide-phase modules (4 of 10), showed the largest absolute gains. Advanced users, who start in Collaborator mode (1/3/5/1 distribution), still improved on advanced features like hooks and skills, even though their foundational scores were already near ceiling. This suggests the curriculum provides value across experience levels, though through different mechanisms: foundational scaffolding for beginners, advanced feature depth for experienced users.

\paragraph{Limitations of this evaluation.} This pilot has several constraints that temper interpretation. The study uses self-reported efficacy, not task-based performance; participants may overestimate or underestimate their abilities. There is no control group, so gains could partially reflect practice effects or exposure rather than the specific pedagogical design. The sample is drawn from a single client organization, limiting generalizability. Because participants knew the author designed the curriculum, demand effects on self-report measures are possible, though self-administered anonymous surveys mitigate this risk. Finally, while the 5-point scale is standard, a 7-point Likert scale would provide finer discrimination in future studies.

\section{Discussion}
\label{sec:discussion}

\subsection{Framing as a System Paper}

This work follows the system paper tradition in AI research, where several influential projects have described architectural contributions without user studies. LangChain~\citep{chase2022langchain} documents a composability framework; MetaGPT~\citep{hong2024metagpt} presents a multi-agent collaboration framework. Both describe design decisions, architectural patterns, and capabilities without empirical user evaluation. Our contributions are similarly system-level design decisions: persona progression, unified curricula, step-pacing, and auto-updating design. The pilot evaluation (Section~\ref{sec:evaluation}) provides initial empirical support, with significant reported self-efficacy gains across all 10 assessed skill areas ($n = 27$, all $p < 0.001$). The system is open-source, enabling independent replication and evaluation, and the test suite provides verifiable structural guarantees.

\subsection{Structured Scaffolding, Not Structured Learning}

A clarification on terminology: \textsc{cc-self-train} is not structured learning in the traditional sense of linear courseware with predetermined interactions. The module sequence and feature progression are fixed, but the learning experience within each module is open-ended. The learner converses with an agentic AI that adapts to their questions, debugs their real errors, and responds to the specific state of their project. Two learners completing the same module will have substantively different interactions. The BYOP (Bring Your Own Project) path makes this especially clear: there are no pre-scripted exercises at all; the learner must identify where each Claude Code feature applies within their own codebase. The curriculum is better understood as \emph{sequenced scaffolding around an exploratory agent interaction}, more akin to a choose-your-own-adventure with guided checkpoints than to a traditional self-paced course.

\subsection{Comparison with Existing Resources}

Table~\ref{tab:comparison} compares \textsc{cc-self-train} with existing AI coding assistant learning resources across six dimensions.

\begin{table}[ht]
\centering
\caption{Comparison with existing AI coding assistant learning resources. ``Docs'' refers to vendor reference documentation (e.g., API docs, CLI reference), not vendor-hosted courses, which fall under Bootcamps.}
\label{tab:comparison}
\small
\begin{tabularx}{\textwidth}{l c c c c}
\toprule
\textbf{Dimension} & \textbf{Docs} & \textbf{YouTube} & \textbf{Bootcamps} & \textbf{cc-self-train} \\
\midrule
Progressive structure & Minimal & Rare & Partial & Full (10 modules) \\
Hands-on projects & No & Demo only & Exercises & Full projects \\
Instructional tone & Static & Presenter style & Instructor style & 4-stage progression \\
Feature coverage & Complete & Selective & Selective & Complete \\
Cross-domain transfer & N/A & Single domain & Single domain & 5 domains \\
Adaptive scaffolding & No & No & Rare & Runtime-adaptive \\
Self-updating & Manual & Never & Rarely & Auto-updating \\
\bottomrule
\end{tabularx}
\end{table}

\subsection{Adoption as a Learning Problem}

Industry surveys consistently identify adoption, not access or model capability, as the bottleneck to AI value capture. The MIT NANDA \emph{State of AI in Business 2025} report finds that approximately 95\% of enterprise generative-AI pilots fail to produce operational or financial impact and attributes the gap to a learning failure on the system side: deployed AI does not retain feedback, adapt to context, or improve over time, and the same users who succeed with AI in personal workflows describe enterprise deployments as unreliable~\citep{challapally2025genai}. \textsc{cc-self-train} addresses an inverted form of the same gap. Rather than waiting for the AI to learn the user, we instrument the user's environment so that learning is structured, observable, and adaptive at the \emph{learner} level. The persona progression encodes pedagogical handoff, the engagement classifier converts conversational behavior into a scaffolding signal, and the asymmetric adaptation rule ensures the system responds faster to struggle than to mastery. Two design choices map onto the capabilities NANDA's respondents most often request from successful enterprise AI: systems that learn from feedback (cited by 66\% of executives in the report) and systems that retain context (63\%). Engagement signals written to \mbox{\texttt{learner-profile.json}} provide local feedback memory, and the cross-session persona state retains pedagogical context across module boundaries. The pedagogical implication is that the learning mechanisms NANDA finds missing in deployed AI systems can be supplied around them, by treating tool adoption as an instructional problem rather than a procurement one.

\subsection{Limitations}

\paragraph{Preliminary empirical evaluation.} The pilot study (Section~\ref{sec:evaluation}) provides initial evidence of self-efficacy gains but uses self-report measures rather than task-based performance assessment, lacks a control group, and draws from a single organization. Whether the persona progression produces measurable learning gains compared to a static persona, or whether STOP blocks reduce cognitive load versus unconstrained delivery, remain open questions that require controlled experiments.

\paragraph{Heuristic engagement classification.} The adaptive learning system (Section~\ref{sec:adaptive}) classifies learner engagement using keyword pattern matching rather than LLM-based semantic analysis. This approach avoids per-turn inference costs but may misclassify messages: a student pasting an error message for context (not debugging) would be classified as \texttt{debug\_attempt}, and rhetorical questions could be scored as concept questions. The streak detection window (three consecutive same-type interactions out of a five-element buffer) and the promotion/demotion thresholds (3.8 and 2.0, respectively) were set based on authorial judgment rather than empirical calibration. A misclassified interaction within the streak window could trigger or suppress an intervention. Whether the adaptive system, including the two-timescale design, produces measurable learning gains compared to the static persona schedules remains an open empirical question.

\paragraph{Sync pipeline accuracy.} The sync pipeline (Section~\ref{sec:staying-current}) verifies structural integrity (step numbering, checkpoint preservation, STOP block presence) through the test suite, but does not verify the \emph{semantic} correctness of LLM-generated updates. Whether generated teaching content accurately reflects upstream feature changes, and how often the pipeline succeeds or silently produces incorrect content, has not been empirically measured. In practice, a human maintainer reviews standalone bulk updates before commit, which mitigates the risk for the primary update mode. Onboarding-mode updates are more automated and would benefit from explicit accuracy measurement across multiple release cycles.

\paragraph{Single tool implementation.} The current implementation targets Claude Code. Adapting the framework to other agentic tools (Copilot's agent mode, Cursor's agentic features, or future tools) would require new module content and feature mappings, though the pedagogical architecture, including persona progression, unified curricula, step-pacing, and the auto-updating design pattern, is designed to transfer. More broadly, while the pedagogical patterns (persona progression, step-pacing, auto-updating) are domain-agnostic, all validation in this paper targets software development; applying them to non-development domains such as legal AI tools or scientific research agents remains a direction for future work.

\paragraph{English only.} All 50 module files, 22 context documents, and the onboarding flow are written in English. Localization would require not just translation but cultural adaptation of the persona progression model.

\paragraph{No telemetry.} The adaptive learning system writes engagement signals to \texttt{learner-profile.json}; this local file never leaves the learner's machine. The system collects no centralized usage data. We cannot report aggregate completion rates, common failure points, or time-on-task metrics. The local-only, privacy-respecting design is intentional but limits ongoing empirical evaluation beyond the pilot study reported in Section~\ref{sec:evaluation}.

\subsection{Future Work}

The current system validates the pedagogical architecture but leaves several empirical and engineering questions open. The following extensions would address the most significant gaps:

\begin{itemize}[leftmargin=2em]
  \item \textbf{Semantic engagement classification:} Replace the keyword-based heuristic classifier in the observation hook with an LLM-based or embedding-based classifier that can distinguish genuine concept questions from rhetorical ones and detect engagement patterns invisible to keyword matching. Calibrate the quality scores and adaptation thresholds empirically against learning outcome data.
  \item \textbf{Multi-tool implementations:} Implement the curriculum framework for other agentic coding tools (Copilot's agent mode, Cursor's agentic features), validating that the pedagogical architecture transfers as designed.
  \item \textbf{User telemetry:} With appropriate consent, collect anonymized usage data to measure completion rates, identify drop-off points, and validate the persona progression's impact on learning outcomes.
  \item \textbf{LLM pedagogical evaluation:} Use a separate LLM to evaluate the quality of generated instruction, creating an automated proxy for human expert review.
  \item \textbf{Community extensions:} Enable community-contributed project paths, expanding the cross-domain coverage beyond the current five options.
  \item \textbf{Affect-aware adaptation:} Extend the engagement classifier beyond cognitive markers such as concept questions and debug attempts to capture affective signals such as enjoyment, curiosity, and frustration. Voluntary adoption of AI tools likely depends on affective experience alongside competence; the local profile and the asymmetric adaptation rule could be extended to respond to disengagement as readily as to struggle, and the survey instrument could be augmented with a brief enjoyment item to enable correlation with self-efficacy gains.
\end{itemize}

\subsection{Broader Impact}

Auto-updating curricula introduce a tradeoff between currency and stability: if the sync pipeline propagates an error, all new learners receive incorrect instruction until the maintainer intervenes. The safe-append design (Section~\ref{sec:staying-current}) and the test suite (Section~\ref{sec:qa}) mitigate this risk, but do not eliminate it. Additionally, the curriculum is English-only and assumes access to Claude Code, which requires a paid subscription, potentially limiting access for learners in resource-constrained settings. We believe the open-source release partially addresses this concern by enabling community adaptation and localization.

\section{Conclusion}
\label{sec:conclusion}

We have presented \textsc{cc-self-train}, a modular interactive curriculum framework for learning AI coding assistants through hands-on project construction. Taken together, the five contributions: persona progression, adaptive learning, unified cross-domain sequencing, step-pacing, and auto-updating design, demonstrate that structured, progressive pedagogy for agentic AI tools is both feasible and amenable to automated maintenance. A parametrized test suite enforces structural consistency as a proxy for pedagogical invariants across all 50 module files, providing a quality assurance layer that keeps the curriculum coherent as it evolves.

A pilot evaluation with 27 participants provides initial empirical support: all 10 assessed skill areas showed statistically significant reported self-efficacy gains ($p < 0.001$), with the largest effects on advanced features like hooks and skills ($d$ up to 2.79). Beginners gained the most overall (+1.18 on a 5-point scale), while advanced users still improved on compositional features, suggesting the experience-stratified persona boundaries provide value across skill levels.

Beyond the specific curriculum, five design patterns in \textsc{cc-self-train} are separable from the implementation and applicable to other AI-mediated instructional systems:

\begin{itemize}[leftmargin=2em]
  \item The \emph{staged-persona adaptation} of Gradual Release of Responsibility, reinterpreting the classroom handoff phases (``I do it'' through ``You do it alone'') for single-learner AI instruction.
  \item The \emph{two-timescale engagement adaptation} architecture, combining slow module-boundary adjustments with fast streak-based interventions, governed by an asymmetric response principle that scaffolds struggling learners more readily than it withdraws support.
  \item The \emph{safe-append rule} for updating instructional content without invalidating mid-flow learners, enabling curricula to evolve without breaking in-progress instruction.
  \item The \emph{parametrized structural test suite} as a proxy for pedagogical invariants, making it infeasible for content updates (whether human or LLM-generated) to silently violate pedagogical properties.
  \item The \emph{unified cross-domain curriculum} in which multiple project paths share identical feature sequencing, producing transfer-friendly learning without duplicating pedagogical design effort.
\end{itemize}

More broadly, as agentic AI tools proliferate across domains beyond software development, from legal research to scientific analysis to creative production, pedagogical systems that adapt their scaffolding and update their content automatically will become a necessary complement to tool documentation. The patterns above offer a starting point for that broader design space.

\paragraph{Availability.} The \textsc{cc-self-train} curriculum, test suite, and onboarding agent are open-source and available at \url{https://github.com/zainnab-sparq/cc-self-train} under the MIT License. This paper describes the system at tag \texttt{v2.25.0} (commit \texttt{5f4c7d1}); subsequent versions may differ in implementation details. The repository includes all module files, context documents, test fixtures, and instructions for running the curriculum or invoking the sync pipeline independently. Anonymized pre/post response data from the pilot evaluation (Section~\ref{sec:evaluation}) is available from the corresponding author on request.

\section*{Acknowledgments}

The design of \textsc{cc-self-train} was inspired in part by two open-source projects. Everything Claude Code~\citep{affaan2025ecc} demonstrated that Claude Code's extensibility surface (agents, skills, hooks, and commands) could be systematically organized into a cohesive configuration system, motivating our effort to build a structured learning pathway through those same features. OpenClaw~\citep{openclaw2025} showed how an open-source agent platform with extensible skills could serve a broad developer community, reinforcing our commitment to open, community-accessible educational tooling. We are grateful to the maintainers of both projects for their contributions to the ecosystem.

\bibliographystyle{plainnat}
\bibliography{references}

\appendix
\section{Survey Instrument}
\label{app:survey}

The following 10 items were administered identically as pre- and post-training self-assessments. Each item used a 5-point Likert scale: 1 = Strongly Disagree, 2 = Disagree, 3 = Neither Agree nor Disagree, 4 = Agree, 5 = Strongly Agree. A screening question (``Have you taken the Claude Code training?'') routed participants to the pre- or post-survey.

\begin{enumerate}[leftmargin=2em]
  \item I can explain what Claude Code is and when it is useful.
  \item I can get Claude Code working in a new repository and complete a small change end-to-end (make edits, verify, and finish) without step-by-step help.
  \item I understand what project memory (CLAUDE.md) is in Claude Code.
  \item I can use project memory (CLAUDE.md) to make Claude behave consistently across sessions in the same repo.
  \item I know when to plan a task first (before making edits) versus when to execute immediately.
  \item I understand what a skill (custom slash command) is in Claude Code.
  \item I can create and use a skill (custom slash command) to standardize a repeated workflow.
  \item I understand what a hook is in Claude Code.
  \item I can set up a hook so an automated action runs at the right time during a Claude Code workflow (for example formatting, linting, or tests).
  \item When Claude is off-track, I can troubleshoot effectively by changing context, instructions, or memory to fix the root cause.
\end{enumerate}

\end{document}